# Direct observation of room-temperature exciton condensation

Jiaxin Yu[1,2,*], Guangyu Dai[1,2], Shuai Xing[1,2], Weiwei Zhang[1], Lin Dou[1], Tianci Shen[1], Xinyu Zhang[1], Xialian Feng[1] & Fuxing Gu[1,*]

[1]Laboratory of Integrated Opto-Mechanics and Electronics, School of Optical-Electrical and Computer Engineering, University of Shanghai for Science and Technology, Shanghai, China.

[2]These authors contributed equally: Jiaxin Yu, Guangyu Dai, Shuai Xing.

* Corresponding authors, e-mail: yujiaxin@usst.edu.cn (J. Y.); gufuxing@usst.edu.cn (F. G.).

**Abstract:**

**Exciton condensation—an interaction-driven, macroscopically coherent paired-fermion state—offers the prospect for dissipationless energy transport in solids, akin to that in superconductivity. Although their light effective mass and strong Coulomb binding favour high transition temperatures, convincing demonstrations of pure-exciton condensation have hitherto been limited to cryogenic conditions. Here, we report the direct observation of quasi-equilibrium condensation of dark excitons in monolayer tungsten diselenide at 300 K and ambient pressure. We achieve this by creating nanoscale spacing-graded Stark traps to confine free excitons, setting the finite-size scale, non-resonant off-axis optical injection to control the local density-temperature trajectory, and employing surface plasmon polariton-enhanced microsphere-assisted microscopy to boost dark-exciton emission and directly image first-order spatial coherence with sub-diffraction resolution. We observe a sharp degeneracy threshold and a clear phase transition, evidenced by extended first-order spatial coherence with algebraic decay and a critical exponent consistent with the universal Berezinskii-Kosterlitz-Thouless criterion. Identical condensation signatures are observed in over 30 independent samples. Our work establishes a room-temperature excitonic platform for exploring strongly correlated many-body physics and advancing near-dissipationless, coherent quantum technologies.**

Emergent phenomena, where collective interactions create novel macroscopic properties, provide a unifying framework across modern science[1]. When interactions among quantum degrees of freedom dominate, emergence becomes particularly profound and elusive, marking the strongly correlated many-body regime. Electron platforms—exemplified by high-temperature superconductors—have offered a uniquely fertile ground for this frontier, where long-range phase coherence transforms the hidden complexity of correlation-driven emergent phenomena into quantitative, falsifiable macroscopic benchmarks (e.g., zero resistance and the Meissner effect)[2]. Yet these realisations occur under cryogenic conditions, limiting broad accessibility and scalability. Extending such emergence to ambient conditions is therefore a central frontier: not merely to lower experimental barriers, but to establish clean, controllable, and information-rich platforms to explore strongly correlated many-body physics in the warm, noisy environments where life and technology operate—thereby opening pathways spanning ambient-condition superconductivity and the quantum–classical boundary[3,4]. However, existing room-temperature platforms remain fragmented: strongly correlated electron platforms (e.g., strange metals) lack macroscopic coherence[5], whereas coherent platforms (e.g., polaritons and magnons) are typically weakly interacting[6,7]. This motivates a straightforward question: Can we realise a single emergent platform that unites strong correlations with robust macroscopic coherence—akin to high-temperature superconductors—while operating under ambient conditions?

Here, we answer this question by reporting the quasi-equilibrium condensation of excitons—bosons with strong Coulomb interactions—at 300 K and ambient pressure. Excitons in two-dimensional (2D) transition-metal dichalcogenides (TMDs) are natural candidates[8–10]: their small effective mass favours high-temperature degeneracy, while large Coulomb binding energies ensure robustness. Beyond electrical methods, excitons enable direct, real-time, and non-invasive optical access to their phase, momentum, and energy properties—providing rich observables of many-body orders. Their control manifold (e.g., dielectric screening, electrostatic displacement field, and strain engineering) also enables Hamiltonian engineering in clean, highly tunable settings for strongly correlated physics. Historically, a lifetime–binding trade-off has limited condensation temperatures below ~100 K[11]: long-lived excitons (e.g., interlayer) are often weakly bound[12,13], whereas strongly bound ones (e.g., bright intralayer) decay too rapidly. Spin-forbidden dark excitons mitigate this dilemma yet are optically inactive[14]. Despite progress in high-temperature excitonic insulators and moiré excitons[15–17], unambiguous macroscopic coherence at room temperature has remained elusive.

To overcome these challenges, we built an experimental system that injects, manipulates, and detects dark excitons in pristine monolayer tungsten diselenide ($WSe_2$). It integrates three key elements: nanoscale spacing-graded Stark traps to tune exciton-exciton interactions and set the finite-size scale, non-resonant off-axis optical injection to shape the local density-temperature trajectory, and a surface plasmon polariton (SPP)-enhanced microsphere-assisted microscopy (MAM) to boost dark-exciton emission and directly image first-order spatial coherence with sub-diffraction resolution. We observe a sharp quantum-degeneracy threshold and a clear phase transition; crucially, the directly imaged spatial coherence shows algebraic decay with an exponent approaching the universal Berezinskii-Kosterlitz-Thouless (BKT) critical value, indicating the onset of interaction-dominant phase stiffness[18,19]. Together, these results constitute the direct observation of a 2D BKT-type exciton condensate.

## Spacing-graded Stark trap

Figure 1a schematically depicts our trap-fabrication concept that confines free excitons without physical contacts. A vertical potential difference $\Delta V_{ms}$ exists intrinsically between a monolayer semiconductor and a metallic substrate due to their work-function difference $\phi_{mat} - \phi_{sub}$, giving rise to a local electric field $E(z) \propto \Delta V_{ms}/z$ at a vertical separation $z$. Instead of tuning $E$ by varying $\Delta V_{ms}$, we exploit a template-assisted thermal morphology control (TA-TMC, Extended Data Fig. 1) process to continuously modulate $z$ on the nanoscale (see Methods). The resulting inhomogeneous electric field causes a second-order Stark shift[20], yielding a potential $U = -\tfrac{1}{2}\alpha|E|^2$ that confines excitons in a trap centred at the minimum of $z$ ($\alpha$ is the exciton polarizability).

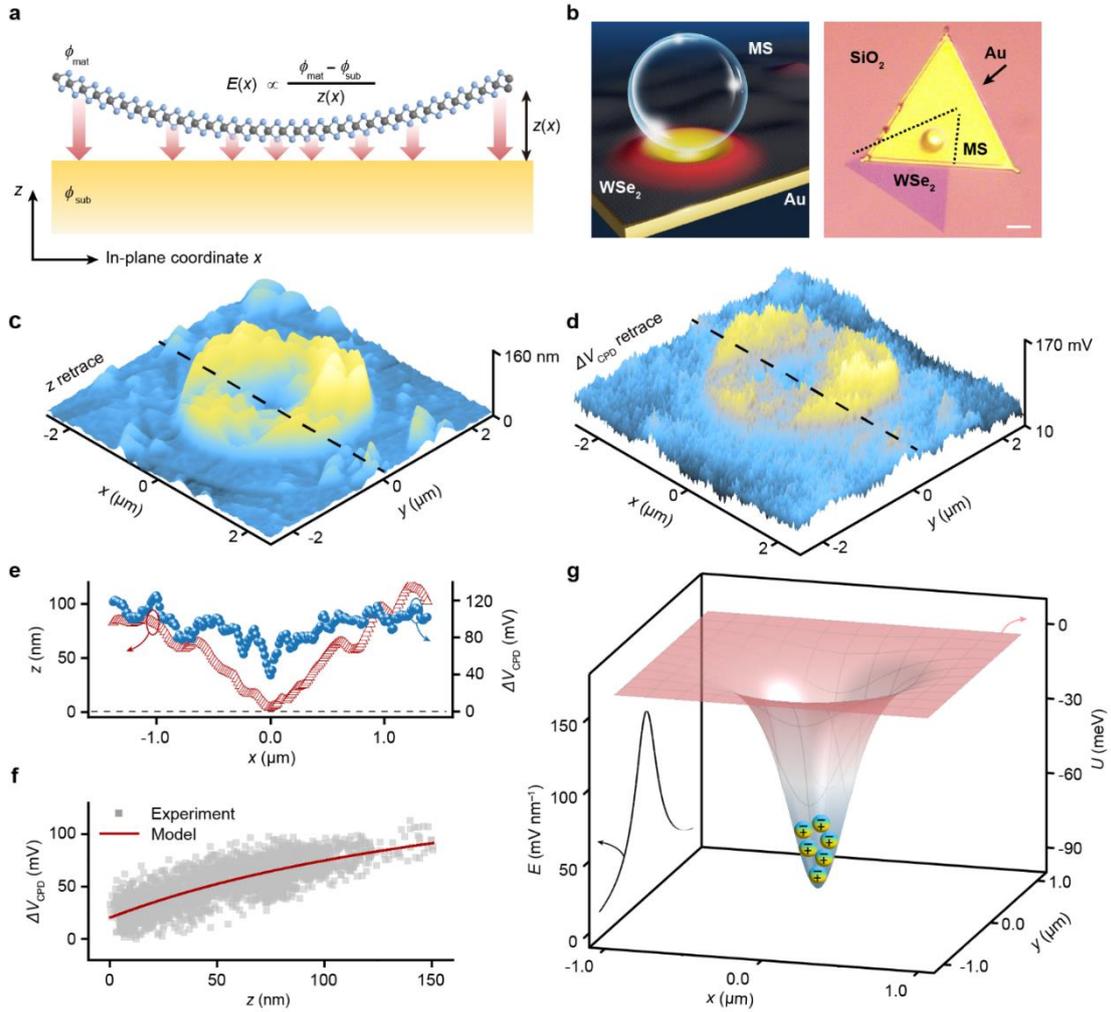

**Fig. 1 | Spacing-graded Stark trap for excitons. a**, Schematic of the potential trap. A curved monolayer suspended above a planar metal substrate creates a varying spacing $z$. The work-function difference between monolayer ($\phi_{mat}$) and substrate ($\phi_{sub}$) generates a position-dependent out-of-plane electric field $E$ that confines excitons. **b**, Fabrication concept using TA-TMC (left) with a microsphere (MS) as the template and an optical micrograph of a typical device layout (right). Scale bar, 5 μm. **c,d**, AFM topography (**c**) and the corresponding map of $\Delta V_{CPD}$ (**d**) for a typical fabricated trap. **e**, Line profiles of $z$ (open triangles) and $\Delta V_{CPD}$ (filled circles), taken along the cross-sections indicated in **c** and **d**, respectively. The dashed line marks $z = 0$. **f**, $\Delta V_{CPD}$ versus $z$ with global fit to the three-capacitor

coupling model. **g**, Reconstructed 3D exciton confinement potential $U$ derived from $E$ (side panel: central cut at $x =$ 0) using local-fit parameters extracted near the trap bottom ($z < 10$ nm). The potential depth is about 90 meV, well above $k_BT$ at room temperature.

In TA-TMC, a low-thermal-conductivity silica ($SiO_2$) microsphere placed on a wrinkled monolayer $WSe_2$ atop a single-crystal Au substrate breaks in-plane thermal symmetry during annealing (Fig. 1b). Differential expansion conforms the monolayer onto the sphere, forming a microscale caldera-like suspended structure with spacing graded from 1 nm at the centre to 100 nm at the edge. Figure 1c shows an atomic-force microscopy (AFM) image of a representative trap formed with an 8-µm-diameter $SiO_2$ sphere, revealing the expected caldera morphology with a lateral diameter of < 2 µm. The monolayer shows an apparent topographic height of $1.5 \pm 0.5$ nm at the trap centre—well above its intrinsic thickness—suggesting genuine suspension (Extended Data Fig. 2). Additionally, the interior surface exhibits atomic flatness, whereas the exterior regions inherit the substrate's high roughness (Extended Data Fig. 3). Spatially photoluminescence (PL) and uniform Raman mapping (Extended Data Fig. 4) further exclude direct Au contact. Unlike patterned electrostatic or strain-induced traps with exciton localisation, quantum confinement or even ionization[21,22], our contact-free architecture avoids such drawbacks, facilitating free exciton formation while preserving their large binding energy.

The vertical electric field $E$ is derived from the AFM-measured spacing map $z$ and the scanning Kelvin-probe microscopy (SKPM)-mapped relative contact potential difference $\Delta V_{CPD}$ between the material and a bare Au substrate (Fig. 1d, see Methods). A typical cross-section (Fig. 1e) shows a clear spatial correlation between $\Delta V_{CPD}(x)$ and $z(x)$, and statistics from multiple profiles (Fig. 1f) are captured by a three-capacitor coupling model (Extended Data Fig. 2, see Supplementary Information). Local fits of the trap bottom ($z < 10$ nm) across devices reconstruct a centre-to-edge field difference of approximately 120 mV nm$^{-1}$, producing a $U \approx 90$ meV confining potential (Fig. 1g)—well above $k_BT$ at 300 K (26 meV). Such a deep potential strongly suppresses thermal escape and enables exciton accumulation. Approximating $U$ near its minimum as a 2D harmonic oscillator yields a characteristic length of about 10 nm ($\gg$ the 2–3 nm exciton Bohr radius) that avoids strong quantum confinement, and an energy level spacing of 1 meV that is small enough to allow the exciton gas to be treated as continuous.

**Dark-exciton injection and detection**

Far-field mapping by optical microscopy (OM) shows that the emission at the trap edge is much brighter than at the centre (Fig. 2a). Yet the photon energy and linewidth are identical at both positions and coincide with the free-exciton absorption peak at 1.669 eV (Extended Data Fig. 4), excluding strain-induced localisation or funnelling inside the trap. The emission energy also exceeds that of localised excitons outside the trap, corroborating that the signal originates from free, bright excitons. In monolayer TMDs, bright excitons carry an in-plane transition dipole moment, couple efficiently to out-of-plane radiation (optical emission wavevector k) and are readily detected by far-field OM, whereas dark excitons possess an out-of-plane moment, couple mainly to high-$k$ in-plane fields and are therefore far-field silent[23] (Fig. 2b). In WSe$_2$, the dark state lies tens of millielectronvolts below the bright state, and is consequently expected to host a non-negligible thermal population at 300 K.

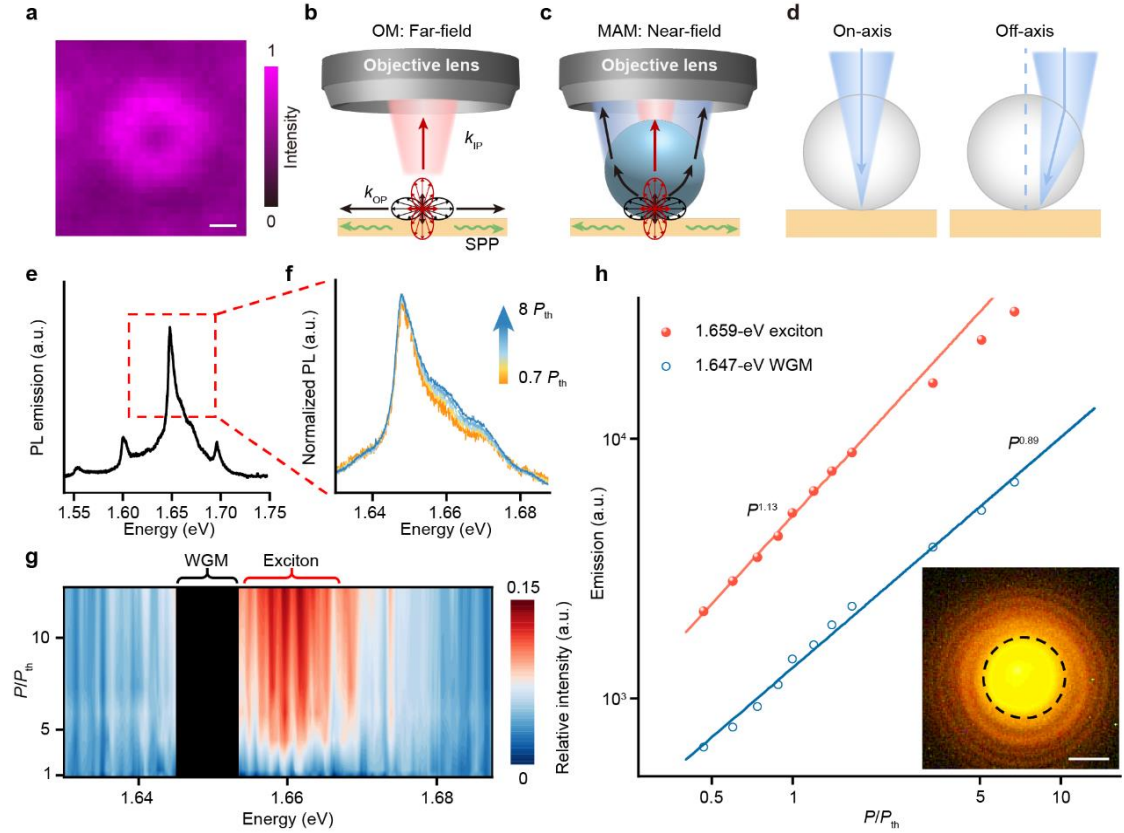

**Fig. 2 | Exciton injection and detection. a**, PL intensity map of a fabricated exciton trap, showing suppressed emission at the centre. **b,c**, Schematics comparing collection efficiency for dark-exciton emission with an out-of-plane dipole ($k_{OP}$, black). In contrast to bright-exciton emission with an in-plane dipole ($k_{IP}$, red), $k_{OP}$ couples weakly to OM (**b**) because it lies largely outside the objective's numerical aperture (NA), but is efficiently collected by MAM (**c**), where the microsphere raises the effective NA and enables SPP near-field coupling. **d**, Pump geometries relative to the microsphere: on-axis (left) and off-axis (right). **e,f**, MAM-collected PL (**e**). The 1.63–1.69 eV window in **e** is enlarged in **f**, showing normalised spectra for pump power from 0.7 to 8 $P_{th}$ and revealing a new line emerging at 1.659 eV. **g,h**, False-colour spectral evolution of relative PL, defined as $[I(P) - I(P_{th})]/I(P_{th})$, with pump power (**g**). Braces mark integration windows centred at 1.659 eV (exciton) and 1.647 eV (WGM). The corresponding integrated intensities (**h**, log–log) show super-linear growth for the excitonic line (slope ≈ 1.13) and sub-linear growth for the WGM (≈ 0.89). Inset, far-field PL image showing concentric interference fringes. Scale bars, 1 μm (**a**) and 5 μm (**h**).

To detect dark excitons, we employ SPP-enhanced MAM with a 6.5-μm-diameter SiO$_2$ microsphere (Fig. 2c). SPPs on the Au substrate provide a high local density of optical states in-plane channel that couples strongly to the out-of-plane dipole of dark excitons, enhancing their emission[24]. The microsphere folds these high-$k$ in-plane wavevectors towards the optical axis of the microscope, thereby extending the effective spatial-frequency passband beyond the Abbe cutoff by converting near-field components to far-field radiation[25]. Together, these two mechanisms significantly boost the collection efficiency (Extended Data Fig. 10) and achieve sub-diffraction-limited spatial resolution.

For excitation, we use an off-resonant, continuous-wave 1.96 eV laser (0.3 eV above the 1$s$ bright state). Instead of centring the beam on the microsphere axis (on-axis injection), we offset it by 1.5–2 μm to illuminate the trap edge (off-axis injection; Fig. 2d), for reasons discussed in Fig. 4.

**Signatures of exciton condensation**

As the pump power rises from 5 nW to a threshold $P_{th} \approx 100$ nW, whispering gallery modes (WGMs) persist (Fig. 2e), yet a new peak emerges at 1.659 eV, spectrally distinct from the WGMs (Fig. 2f, g), and grows super-linearly with pump power—in contrast to the throughout sub-linear growth of the WGMs lines (Fig. 2h). Furthermore, the peak energy does not track the microsphere resonance shifts (Extended Data Fig. 5), ruling out cavity-coupled modes as the origin.

Far-field PL imaging through the microsphere reveals a pronounced ring-shaped interference pattern (inset of Fig. 2h), evidencing point-like coherence from the exciton ensemble—incompatible with incoherent recombination or WGM leakage (Fig. S2 in Supplementary Information). Two control samples, where SPP enhancement is removed (WSe$_2$ on SiO$_2$) or the dark-exciton population is suppressed (MoSe$_2$ on Au)[27], both eliminate the coherent emission (Extended Data Fig. 6). The comparison confirms the essential role of SPP-coupled emission from a densely populated dark-exciton state in producing the observed coherence. Notably, the 1.659 eV peak lies near the bright-exciton line—rather than at the lower, dark-exciton line[27]—a placement that may be ascribed to a chiral-phonon–assisted up-conversion of dark excitons[28] (Supplementary Information).

The observed threshold-like behaviour and ring-shaped interference are indicative of a macroscopic, phase-coherent dark-exciton state, with phase-space density above its thermal value. The definitive metric, however, is the extension of spatial coherence[29]. Typically, spatial coherence grows rapidly near the quantum-degeneracy threshold and can extend over hundreds of thermal de Broglie wavelengths ($\lambda_T$) in macroscopic condensates (e.g. 100 $\lambda_T \approx 600$ nm at 300 K). This length is comparable to the smallest observable coherence length defined by the point-spread function (PSF) of conventional far-field OM, obscuring any distinction from incoherent states. In contrast, our MAM system, with enhanced effective NA and magnification (Extended Data Fig. 7), compresses the PSF by a factor of 1.92, yielding spatial resolution down to ~50 $\lambda_T$ (Fig. 3a, Extended Data Fig. 8). Combined with a point-inversion Michelson interferometer (Fig. 3b, see Methods), it enables direct measurements of first-order spatial coherence and thus observation of the onset of a macroscopically coherent quantum state.

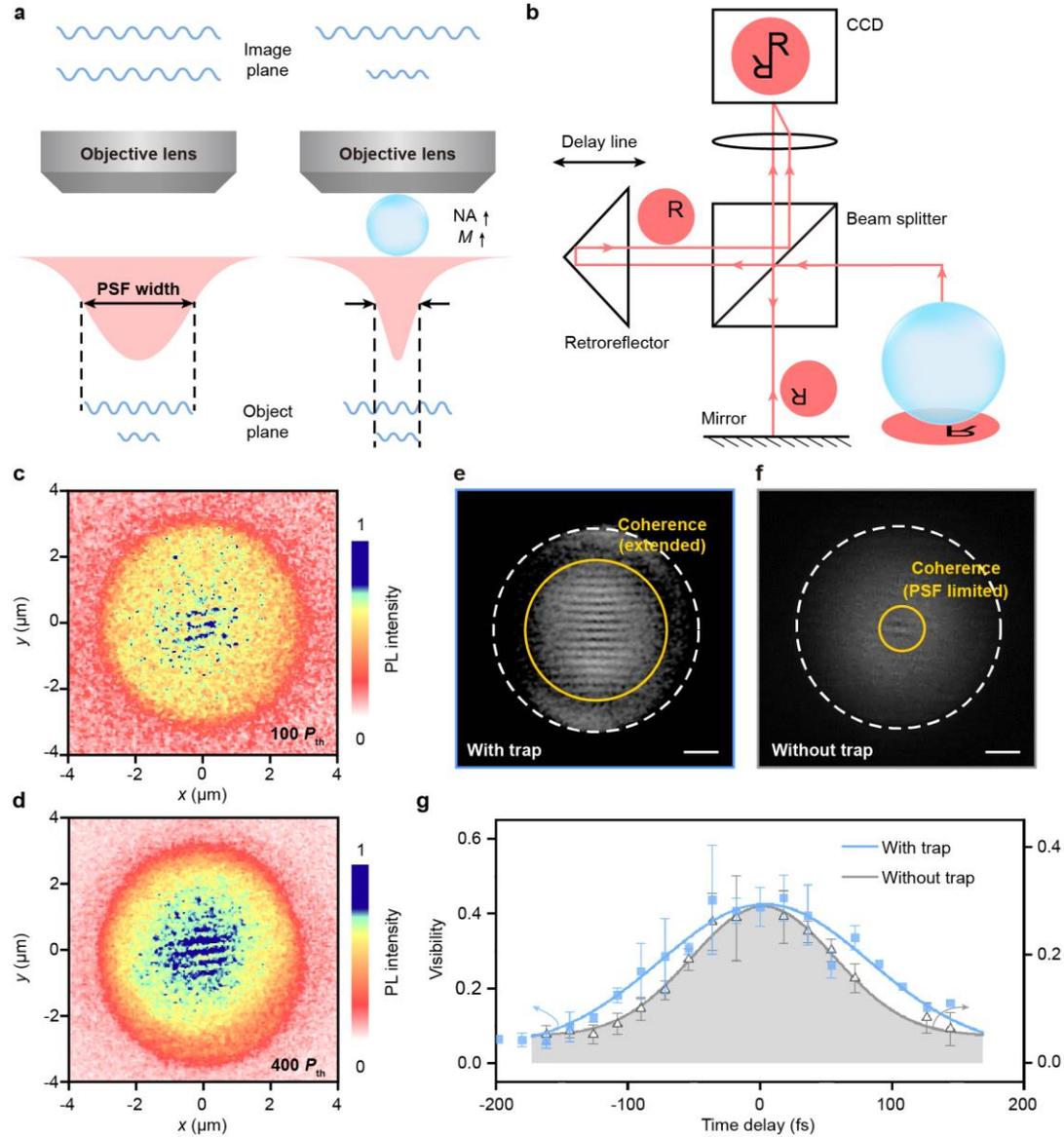

**Fig. 3 | Spatial and temporal coherence. a**, Schematics comparing PSFs for OM (left) and MAM (right). In MAM, the microsphere increases the effective NA and provides magnification (*M*), yielding a smaller object-plane PSF and thereby lowering the minimum detectable spatial-coherence length. **b**, MAM-based point-inversion Michelson interferometer. The sample image and its centrosymmetric replica interfere on the CCD; a motorised delay line controls the temporal delay. **c,d**, Real-space interferograms of the 1.659-eV emission line recorded at zero delay for pump power of 100 $P_{th}$ (**c**) and 400 $P_{th}$ (**d**). **e,f**, Interferograms at a saturation power (of order $10^3$ $P_{th}$). The coherence for a region containing a trap (**e**) extends beyond the PSF, whereas for a region without a trap (**f**) it remains PSF-limited. White dashed circles mark the MAM field of view (FOV). Scale bars, 1 μm. **g**, Fringe visibility versus time

delay for regions with a trap (filled squares) and without a trap (open triangles). The visibility $V = (I_{max} - I_{min})/(I_{max} + I_{min})$ is position-averaged at each delay over multiple locations—outside the PSF (with trap) or inside the PSF (without trap); error bars denote the corresponding standard deviation. Gaussian fits yield coherence times of 159 fs (with trap) and 108 fs (without trap).

We then track the evolution of spatial coherence as a function of pump power. For power below 100 $P_{th}$, the interference area is limited by the PSF (Fig. 3c). As the power increases, the spatial coherent footprint expands well beyond the PSF, reaching twice the PSF width at 400 $P_{th}$ (Fig. 3d), then saturates at a size set by the trap geometry (Fig. 3e), with the phase remaining locked throughout. By contrast, under identical pump power but without the trap, the fringes never extend beyond the PSF (Fig. 3f) and their coherence decays within 108 fs—much faster than the 159 fs in the trapped case (Fig. 3g), a timescale that essentially matches the average exciton–phonon scattering interval at room temperature[30]. These observations highlight both spatial and temporal phase coherence enabled by exciton confinement.

To quantify the phase transition, we extracted the first-order spatial coherence $g^{(1)}$, which is related to the coherence function $g_{ideal}$ of an ideal imaging system by a convolution with the PSF (denoted by $f$) of the real system: $g^{(1)}(\Delta r) = g_{ideal} \otimes f(\Delta r)$, where $\Delta r$ denotes the separation between the positions $(x, y)$ and $(-x, -y)$ in the object plane (see Methods). Under on-axis injection, $g^{(1)}$ decays exponentially at the PSF length scale, as expected for a uniform classical gas (Fig. 4a). Under off-axis injection, $g^{(1)}$ remains sizable over ~250 $\lambda_T$ (Fig. 4b,c), far beyond the PSF. Within a finite-temperature 2D-BKT framework[31], thermal phase-phonon fluctuations give $g_{ideal}(\Delta r) \propto (\Delta r)^{-\eta}$. Fitting the off-axis data yields a critical exponent $\eta = 0.33 \pm 0.06$ (Fig. 4c), approaching the universal BKT criterion $\eta_c = 1/4$ and corresponding to $0.75 \pm 0.13$ of the Nelson–Kosterlitz critical phase-stiffness jump[19]. Occasional fork-like dislocations from vortex–antivortex unbinding are also observed, further supporting a BKT-type condensate of strongly interacting excitons at 300 K.

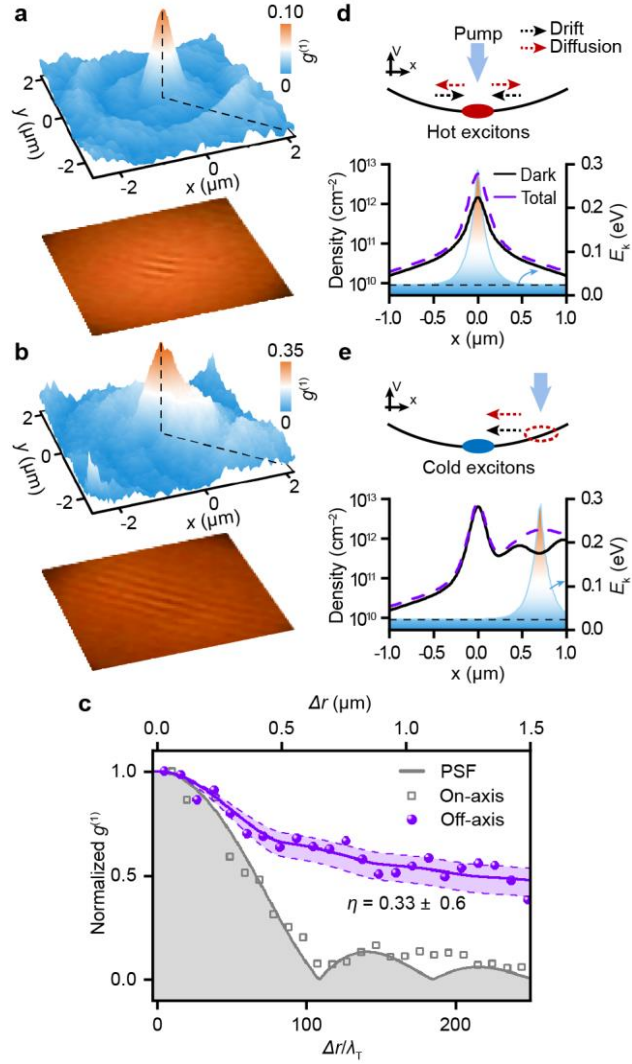

**Fig. 4 | BKT-type condensation enabled by off-axis injection. a,b**, Maps of $g^{(1)}$ (top) and the corresponding interferograms (bottom) under on-axis (**a**) and off-axis (**b**) injection. On-axis injection gives $g^{(1)}(0) \approx 0.10$ that decays to zero within about 0.8 μm on the image plane, whereas off-axis injection yields $g^{(1)}(0) \approx 0.35$ that remains finite out to about 2.8 μm. **c**, Normalised $g^{(1)}$ versus separation $\Delta r$ on object plane. The on-axis data (open squares) follow the PSF envelope (grey shading), whereas the off-axis data (filled circles) exhibit an algebraic decay consistent with a 2D BKT form. The fit gives $\eta = 0.33 \pm 0.06$ (purple band), approaching the BKT critical value 1/4. **d,e**, Schematics and drift-diffusion modelling contrasting exciton dynamics for on-axis (**d**) and off-axis (**e**) injection. On-axis injection produces a hot pile-up at the trap centre (small dark-exciton fraction and $E_k \gg k_B T$; dashed line marks $k_B T$). Off-axis injection drives excitons inward by drift and diffusion, yielding a cold, dense pool at the trap centre (dominated dark-exciton fraction and $E_k \approx k_B T$).

## Discussion and outlook

Taken together with the threshold-like behaviour and the simultaneous spatial and temporal coherence, these results provide direct observation of a 2D quasi-equilibrium, BKT-type condensation of dark excitons under ambient conditions. The condensation emerges only when off-axis pumping is combined with trap confinement (Extended Data Fig. 9), a condition determined by exciton density and rapid phonon cooling (Supplementary Information). Steady-state modelling shows that, even at the saturation power (of order $10^3\ P_{th}$), on-axis pumping or removal of the trap pins the dark-exciton density at about $10^{12}\ cm^{-2}$ via local exciton–exciton annihilation—a value that barely reaches the quantum-degeneracy threshold (Fig. 4d, Extended Data Fig. 10). By contrast, the off-axis pumping with trap configuration drives the dark-exciton density to nearly $10^{13}\ cm^{-2}$—far exceeding the degeneracy threshold—thereby making it possible to meet the density prerequisite for condensation (Fig. 4e).

Strong exciton–phonon coupling and the high phonon occupation at 300 K enable hot excitons to transfer their kinetic energy $E_k$ to the lattice on a sub-picosecond timescale[32]—well within their 150–500 ps lifetime (Supplementary Information). The associated cooling length (10–100 nm) is much shorter than the 500-nm pump spot. Consequently, off-axis injection allows excitons to escape the illuminated region via potential-driven drift and phonon-assisted diffusion, cooling en route and accumulating at the trap centre. With on-axis injection, however, the inward trap potential prevents such escape, leading to hot-phonon reabsorption and thereby suppressing exciton thermalisation.

Condensation also requires a narrow pump-power window of 30–60 μW—comfortably above the quantum-degeneracy threshold yet below the Mott-dissociation limit (exciton density $> 10^{13}\ cm^{-2}$)[33]. Although narrow, this window is readily accessible and reproducible: identical condensation signatures are observed in over 30 independent samples. Despite its narrowness, room-temperature operation presents two distinct benefits. First, bound complexes such as biexcitons are thermally unstable[34] and therefore do not deplete the exciton reservoir. Second, the high phonon population accelerates exciton thermalisation within their lifetime, enabling the gas to reach quasi-equilibrium.

Combining strong interactions with macroscopic coherence, our room-temperature dark-exciton condensate marks a decisive advance toward a new platform for exploring strongly correlated many-body physics and advancing near-dissipationless, coherent quantum technologies. Built on standard, cost-effective 2D semiconductors, it offers broad accessibility, reproducibility, and seamless on-chip integration. Crucially, its macroscopic phase coherence transforms complex many-body correlations into quantitative, falsifiable macroscopic benchmarks—including BKT universal signatures (e.g., phase-stiffness jump and algebraic order)—while and a quantitatively mapping experimental controls (e.g., density and trap curvature) onto running couplings for renormalization group-flow analysis. As a long-lived coherent fluid, the condensate facilitates near-dissipationless energy and information transport, laying the foundation for high-speed, ultra-low-power logic circuits for quantum sensing, simulation, computing, and beyond[9,10,35,36]. Moreover, anchored by its topological (BKT) coherence benchmark, the platform naturally connects to topological quantum matter and devices[37,38], and could also inspire exciton-mediated pathways to ambient-condition superconductivity[3].

**Acknowledgements** We are grateful to S.L. Zhuang for his invaluable support and assistance in our project. We thank Q.W. Zhan, W. Fang (Zhejiang University), and S.J. Zhang for their insightful discussions. We are also grateful to Y. Liang for her contribution to the construction of the experimental apparatus. This work was supported by the AI Promotion Fund for Research Paradigm Reform of Shanghai Municipal Education Commission (grant No. Z-2024), and National Natural Science Foundation of China (grant Nos. 12074259 and 62122054).

**Competing interests** The authors declare no competing interests.

**Supplementary information** The online version contains supplementary material available from the authors